\documentclass[11pt,oneside,english]{amsart}
\usepackage[T1]{fontenc}
\usepackage[latin9]{inputenc}
\usepackage{amssymb}

\makeatletter
\numberwithin{equation}{section} 
\numberwithin{figure}{section} 
  \theoremstyle{remark}
  \newtheorem*{acknowledgement*}{Acknowledgement}

\usepackage{babel}
\makeatother

\begin{document}

\title[Distance in space of elasticity tensors]{Invariant properties for finding distance in space of elasticity tensors}

\author{Ioan Bucataru$^{\clubsuit}$ \and Michael A. Slawinski$^{\spadesuit}$}

\thanks{$^{\clubsuit}$Faculty of Mathematics, \char`\"{}Al. I. Cuza\char`\"{}
University, Iasi 700506, Romania}

\thanks{$^{\spadesuit}$Department of Earth Sciences, Memorial University,
St. John's NL A1B 3X5, Canada}

\email{bucataru@uaic.ro, mslawins@mun.ca}

\urladdr{http://www.math.uaic.ro/\textasciitilde{}bucataru/, http://www.mun.ca/geomechanics/}

\begin{abstract}
Using orthogonal projections, we investigate distance of a given
elasticity tensor to classes of elasticity tensors exhibiting
particular material symmetries. These projections depend on the
orientation of the elasticity tensor, hence the distance is
obtained as the minimization of corresponding expressions with
respect to the action of the orthogonal group. These expressions
are stated in terms of the eigenvalues of both the given tensor
and the projected one. The process of minimization is facilitated
by the fact that, as we prove, the traces of the corresponding
Voigt and dilatation tensors are invariant under these orthogonal
projections. For isotropy, cubic symmetry and transverse isotropy,
we formulate algorithms to find both the orientation and the
eigenvalues of the elasticity tensor that is endowed with a
particular symmetry and is closest to the given elasticity tensor.
\end{abstract}
\maketitle

\section{Introduction}

Computing distance between elasticity tensors is particularly
important in inverse problems. For instance, we wish to know
whether or not a material can be considered as isotropic with its
anisotropic behavior accounted for by the errors of inversion,
which stem from both measurement errors and the fact that the
elasticity tensor itself is an idealization in the realm of
continuum mechanics.

A great portion of seismology, which is the motivation for our
study, is performed under the assumption of isotropy due to its
sufficient accuracy and mathematical convenience. However, to increase
this accuracy, several seismologists formulated their equations
without assuming isotropy; notably, Rudzki \cite{Rudzki1,Rudzki2}
and, almost half a century later, Helbig \cite{Helbig1956}. Three
decades after Helbig, Thomsen \cite{Thomsen} popularized the
concept of anisotropy among exploration seismologists by
formulating a linear approximation whose accuracy stems from the
fact that most natural phenomena investigated in seismology are
only weakly anisotropic.

The purpose of this paper is to investigate the concept of
distance in the space of elasticity tensors between a given tensor
and classes of tensors exhibiting particular material symmetries,
and to obtain the effective elasticity parameters and effective
orientations for such symmetries. To facilitate this nonlinear
process, we propose the use of invariants of the orthogonal
projections onto various symmetry classes of elasticity tensors.

The concept of distance of a given elasticity tensor to a
material-symmetry class was introduced by Gazis et al.
\cite{Gazis} using orthogonal projections of elasticity tensors on
the space of a particular class, and by Fedorov \cite{Fedorov},
who minimized the mean-square difference of the slowness surface.
Three decades later, Norris \cite{Norris06} proved that the two
approaches are equivalent to one another. Moakher and Norris
\cite{Moakher} give a detailed presentation of distance of an
elasticity tensor to all material-symmetry classes. In this work,
we use the Euclidean distance, which is induced by the Frobenius
norm. For different distances with a detailed discussion of their
properties we refer the reader to Moakher \cite{Moakher02} and
Moakher and Norris \cite{Moakher}.

In all approaches, the problem raised by Gazis et al. \cite{Gazis}
remains open: the orthogonal projection, $c^{sym}$, of a given
elasticity tensor, $c$, onto the set of elasticity tensors belonging
to a particular material symmetry class, ${\mathcal Sym}$, depends
on the original orientation of tensor $c$. It follows that the
distance, $d\left(c, c^{sym}\right)$, between an elasticity tensor
and its orthogonal projection depends on the a priori orientation of
tensor $c$. Therefore, the distance, $d\left(c, {\mathcal
Sym}\right)$, between an elasticity tensor, $c$, and some symmetry
class, ${\mathcal Sym}$, can be obtained by finding the orientation
of elasticity tensor $c$ that minimizes the distance $ d\left(c,
c^{sym}\right)$. There have been several attempts to solve this
problem. Gazis et al. \cite{Gazis} give necessary conditions for the
cubic case, such that the axes of a coordinate system coincide with
the axes of its closest cubic tensor. Arts et al. \cite{Arts}
propose a method to find particular reference axes in which most
offdiagonal terms are minimized. Dellinger \cite{Dellinger05}
proposes a numerical method to find the rotation-symmetry axis that
minimizes the expression for distance to transverse isotropy.
François et al. \cite{Francois98} and Norris \cite{Norris06} propose
numerical methods to find the closest projection.

We begin this paper by invoking the elasticity tensor together
with the corresponding Voigt and dilatation tensors. We view
elasticity tensors as fourth-rank tensors in $\mathbb{R}^{3}$ and
consequently look for invariant properties under the action of
orthogonal group $O\left(3\right)$. This point of view was
introduced and investigated independently by Walpole
\cite{Walpole}, Rychlewski \cite{Rychlewski84,Rychlewski95}, and
Cowin and Mehrabadi \cite{Cowin_Mehrabadi,Cowin92}.

For a fixed orthonormal basis in ${\mathbb R}^3$, any other
orthonormal basis can be given in terms of an orthogonal
transformation $A\in O(3)$. The orthogonal projection,
$c^{sym}(A)$, of an elasticity tensor, $c$, to a symmetry class,
${\mathcal Sym}$, depends on the orientation, $A$, that is used to
express the elasticity tensor. These orthogonal projections were
considered first by Gazis et al. \cite{Gazis}, and explicitly
obtained by Moakher and Norris \cite{Moakher} for all symmetry
classes. The corresponding Euclidian distance, $d\left(c,
c^{sym}(A)\right)$, between an elasticity tensor, $c$, and its
orthogonal projection, $c^{sym}(A)$, may not represent the
distance to a symmetry class, $d\left(c, {\mathcal Sym}\right)$,
since $d\left(c, c^{sym}(A)\right)\geq d\left(c, {\mathcal
Sym}\right)$. Therefore, we obtain the distance to a given
symmetry class by minimizing $ d\left(c,c^{sym}(A)\right)$, for
$A\in O(3)$. To facilitate this minimization, we express $
d\left(c,c^{sym}(A)\right)$ as function
$f\left(\lambda_{\alpha}^{sym}(A)\right)$, where
$\lambda_{\alpha}^{sym}(A)$ are the eigenvalues of the orthogonal
projection $c^{sym}(A)$.

In this work we prove that the traces of the associated Voigt and
dilatation tensors, which are orthogonal invariants, also remain
invariant under these orthogonal projections, and therefore
certain relations between eigenvalues are preserved. These
invariant relations are useful in simplifying function
$f\left(\lambda_{\alpha}^{sym}(A)\right)$. Since the minimizing
process is taken over the compact group $O(3)$, the minimum is
reached for an orthogonal transformation, $A_e$, which will be
called the effective transformation or orientation. This allows us
to obtain the distance of a given elasticity tensor to a
particular symmetry class, and to find the orientation and
eigenvalues -- the effective elasticity parameters -- of the
closest elasticity tensor that belongs to that class. We discuss
explicitly distance to isotropy, cubic symmetry and transverse
isotropy, and provide numerical examples for our algorithms.

For the case of isotropy, we show that the orthogonal projection
proposed by Moakher and Norris \cite{Moakher} can also be obtained
using the harmonic decomposition of an elasticity tensor, studied by
Backus \cite{Backus} and Cowin \cite{Cowin}. In the case of
isotropy, the orthogonal projection, $c\mapsto c^{iso}$, is
invariant under the action of the orthogonal group. Consequently,
the eigenvalues of the isotropic projection are orthogonal
invariants and therefore they can be referred to as the effective
isotropic invariants. It implies that $f\left(\lambda_{1}^{iso}(A),
\lambda_{2}^{iso}(A) \right)$ is constant. Therefore any orientation
is an effective orientation, $d\left(c, c^{iso}\right)=d\left(c,
{\mathcal Iso}\right)$, and no minimization algorithm is required.
We express this distance in terms of the effective isotropic
parameters, as well as in terms of traces of Voigt and dilatation
tensors for the isotropic projection.

For the case of cubic symmetry, we use the orthogonal projection
proposed by Gazis et al. \cite{Gazis} and by Moakher and Norris
\cite{Moakher}. For an arbitrary orientation, $A$, of an
elasticity tensor, $c$, its orthogonal projection, $c^{cube}(A)$,
has three distinct eigenvalues, $\lambda_1^{cube}(A)$,
$\lambda_2^{cube}(A)$, $\lambda_3^{cube}(A)$, with multiplicities
$1$, $2$ and $3$, respectively. The invariance of the traces of
the Voigt and dilatation tensors under orthogonal projection,
$c\mapsto c^{cube}$, implies that the first eigenvalue of the
cubic projection is an orthogonal invariant and the other two
eigenvalues satisfy a linear relation, which is also an orthogonal
invariant. Therefore, the distance, $
d\left(c,c^{cube}(A)\right)$, can be expressed as a function of
one eigenvalue only, $f\left(\lambda_{2}^{cube}(A)\right)$, which
the minimizing algorithm has to take into account to obtain
distance $d\left(c, {\mathcal Cube}\right).$

For the case of transverse isotropy, we use the orthogonal
projection proposed by Moakher and Norris \cite{Moakher}. For an
arbitrary orientation $A$ of an elasticity tensor, $c$, its
orthogonal projection, $c^{TI}(A)$, has four distinct eigenvalues,
$\lambda_1^{TI}(A)$, $\lambda_2^{TI}(A)$, $\lambda_3^{TI}(A)$,
$\lambda_4^{TI}(A)$, with multiplicities $1$, $1$, $2$ and $2$,
respectively. The invariance of the trace of Voigt tensor under
the orthogonal projection $c\mapsto c^{TI}$ implies a linear
relation among the four eigenvalues of the transversely isotropic
projection. Therefore, the distance, $ d\left(c,c^{TI}(A)\right)$,
can be expressed as a function of three eigenvalues only,
$f\left(\lambda_{1}^{TI}(A), \lambda_{2}^{TI}(A),
\lambda_{3}^{TI}(A)\right)$, which the minimizing algorithm has to
take into account to obtain distance $d\left(c, {\mathcal
TI}\right).$

At the end of the paper, we illustrate the presented approach
using examples of strongly and weakly anisotropic solids. For
these examples, the algorithm provides the distances to isotropic,
cubic and transverse isotropic symmetry, as well as the effective
orientation and effective elasticity parameters for each of the
considered symmetry classes.

\section{Elasticity tensor}

In this section, we present the notions that allow us to formulate
and discuss distance between a given elasticity tensor and material-symmetry
classes.

The elasticity tensor is a fourth-rank tensor in $\mathbb{R}^{3}$,
which according to Hooke's law relates the stress and strain tensors,\begin{equation}
\sigma_{ij}=c_{ijkl}\varepsilon_{kl}.\label{eq:stress_strain}\end{equation}

In this paper, Roman indices, $i,j,\dots$, run from one to three,
and Greek indices, $\alpha,\beta,\ldots$, from one to six; summation
over repeated indices is implied. Both stress and strain are
symmetric second-rank tensors in $\mathbb{R}^{3}$. We denote the
space of second-rank symmetric tensors in $\mathbb{R}^{3}$ by
$L_{2,s}\left(\mathbb{R}^{3}\right)$; this is a six-dimensional
linear space in which we consider the scalar product given
by\begin{equation}
\varepsilon_{1}\cdot\varepsilon_{2}:=\operatorname{Tr}\left(\varepsilon_{1}\
\varepsilon_{2}^{t}\right),\label{eq:scalar}\end{equation} where
$\operatorname{Tr}$ stands for trace and $t$ for transpose.
Therefore, one can view an elasticity tensor as a linear map $c:\
L_{2,s}\left(\mathbb{R}^{3}\right)\longrightarrow
L_{2,s}\left(\mathbb{R}^{3}\right)$ that is symmetric, \[
c\varepsilon_{1}\cdot\varepsilon_{2}=\varepsilon_{1}\cdot
c\varepsilon_{2},\ \qquad\forall\varepsilon_{1},\varepsilon_{2}\in
L_{2,s}\left(\mathbb{R}^{3}\right),\] and positive-definite, \[
c\varepsilon\cdot\varepsilon>0,\ \qquad\forall\varepsilon\in
L_{2,s}\left(\mathbb{R}^{3}\right),\ \varepsilon\neq0.\]

Since $c$ is a fourth-rank tensor in $\mathbb{R}^{3}$, its
components with respect to orthonormal basis $\left\{ e_{1},\
e_{2},\ e_{3}\right\} $ are\[ c_{ijkl}=c\left(e_{i}\otimes
e_{j}\right)\cdot\left(e_{k}\otimes e_{l}\right),\] which satisfy
the intrinsic symmetries given by $c_{ijkl}=c_{jikl}=c_{klij}.$
For subsequent use, we invoke the Voigt and dilatation tensors,
which are second-rank symmetric tensors in $\mathbb{R}^{3}$, given
by
\begin{equation} V_{ij}=c_{ikjk}\quad\mathrm{and}\quad\
D_{ij}=c_{ijkk},\label{eq:Voigt}\end{equation} and discussed by
Cowin and Mehrabadi \cite{Cowin_Mehrabadi} and Helbig \cite{Helbig}.

Let us consider the group of orthogonal transformations,
$O\left(3\right)$, in the three-dimensional Euclidean space,
$\mathbb{R}^{3}$. Each such transformation is linear and preserves
the scalar product. For a fixed orthonormal basis ${\mathcal
B}=\left\{ e_{1},\ e_{2},\ e_{3}\right\} $ in ${\mathbb R}^3$, any
other orthonormal basis ${\mathcal B}'=\{f_1, f_2, f_3\}$ can be
identified with an orthogonal transformation $A\in O(3)$, such
that $Ae_i=f_i$, for all $i\in \{1,2,3\}$. Also, we will refer to
orthogonal transformation $A$ as an orthonormal basis.

An orthogonal transformation $A\in O\left(3\right)$ acts on a
second-rank symmetric tensor, $\varepsilon\in
L_{2,s}\left(\mathbb{R}^{3}\right)$, as\[ \left(A\circ
\varepsilon\right) \left(u,v\right):=\varepsilon\left(Au,Av\right),\
\forall u,v\in\mathbb{R}^{3}.\] An orthogonal transformation $A\in
O\left(3\right)$ acts on the elasticity tensor as\[ \left(A\ast
c\right)\left(\varepsilon\right)=A\circ \left(c\left(A\circ
\varepsilon\right)\right),\ \forall\varepsilon\in
L_{2,s}\left(\mathbb{R}^{3}\right).\] For the orthonormal basis
${\mathcal B}=\left\{ e_{1},\ e_{2},\ e_{3}\right\} $, the
components of $A\ast c$ are \[ \left(A\ast
c\right)_{ijkl}=c\left(Ae_{i}\otimes
Ae_{j}\right)\cdot\left(Ae_{k}\otimes Ae_{l}\right).\] We remark
here that these are also the components of elasticity tensor $c$
with respect to the orthonormal basis $A$.

The orthonormal basis $\left\{ e_{1},e_{2},e_{3}\right\} $ of
$\mathbb{R}^{3}$ induces an orthonormal basis of
$L_{2,s}\left(\mathbb{R}^{3}\right)$, given by \begin{equation}
\varepsilon_{\alpha\left(i,j\right)}=2^{-\frac{1}{2-\delta_{ij}}}\left(e_{i}\otimes
e_{j}+e_{j}\otimes
e_{i}\right),\label{eq:ortogonal_omega}\end{equation} where
\begin{equation} \alpha:\left\{ \left(i,j\right),1\leq
i<j\leq3\right\} \longrightarrow\left\{ 1,2,\ldots,6\right\} ,\
\label{eq:alpha}\end{equation} with
$\alpha\left(i,j\right)=i\delta_{ij}+\left(1-\delta_{ij}\right)\left(9-i-j\right)$
and $\delta_{ij}$ being the Kronecker delta. The components
$\varepsilon_{ij}=\varepsilon\left(e_{i},e_{j}\right)$ of a
second-rank symmetric tensor with respect to the orthonormal basis
$\left\{ e_{1},e_{2},e_{3}\right\} $ of $\mathbb{R}^{3}$ can be
arranged into a six-dimensional vector, \[
\left(\varepsilon_{11},\varepsilon_{22},\varepsilon_{33},\sqrt{2}\varepsilon_{23},\sqrt{2}\varepsilon_{13},\sqrt{2}\varepsilon_{12}\right),\]
with respect to basis \eqref{eq:ortogonal_omega}. Since Lord Kelvin
\cite[p. 110]{Kelvin,Kelvin:Vol3} suggested such an approach ---
albeit without using the tensorial notation, which was not known at
the time -- we refer to entities expressed with respect to basis
\eqref{eq:ortogonal_omega} as Kelvin's notation. An elasticity
tensor, which is a second-rank positive-definite symmetric tensor in
$L_{2,s}\left(\mathbb{R}^{3}\right)$, has the components with
respect to basis \eqref{eq:ortogonal_omega} given by\begin{equation}
C_{\alpha\beta}=C\varepsilon_{\alpha}\cdot\varepsilon_{\beta}.\label{eq:cijkl}\end{equation}

$C_{\alpha\beta}$ are the entries of a $6\times6$ matrix, which
we write explicitly as\begin{equation}
\left[\begin{array}{cccccc}
c_{1111} & c_{1122} & c_{1133} & \sqrt{2}c_{1123} & \sqrt{2}c_{1113} & \sqrt{2}c_{1112}\\
c_{1122} & c_{2222} & c_{2233} & \sqrt{2}c_{2223} & \sqrt{2}c_{2213} & \sqrt{2}c_{2212}\\
c_{1133} & c_{2233} & c_{3333} & \sqrt{2}c_{3323} & \sqrt{2}c_{3313} & \sqrt{2}c_{3312}\\
\sqrt{2}c_{1123} & \sqrt{2}c_{2223} & \sqrt{2}c_{3323} & 2c_{2323} & 2c_{2313} & 2c_{2312}\\
\sqrt{2}c_{1113} & \sqrt{2}c_{2213} & \sqrt{2}c_{3313} & 2c_{2313} & 2c_{1313} & 2c_{1312}\\
\sqrt{2}c_{1112} & \sqrt{2}c_{2212} & \sqrt{2}c_{3312} & 2c_{2312}
& 2c_{1312} &
2c_{1212}\end{array}\right].\label{eq:Chapman}\end{equation} This
formulation was introduced and investigated independently by
Walpole \cite{Walpole}, Rychlewski
\cite{Rychlewski84,Rychlewski95}, and Cowin and Mehrabadi
\cite{Cowin_Mehrabadi,Cowin92}. We refer to matrix
\eqref{eq:Chapman}, as Kelvin's notation \cite[p.
110]{Kelvin:Vol3}. This notation has been used by several
researchers; notably, Fedorov \cite{Fedorov}, Helbig
\cite{Helbig}, Chapman \cite{Chapman} and Bóna et al. \cite{BBS1}.

For subsequent use, we invoke the scalar product given
by\begin{equation} c\cdot
c':=c_{ijkl}c'_{ijkl}=C_{\alpha\beta}C'_{\alpha\beta},\label{eq:scalar_el}\end{equation}
which results in the Frobenius norm given by\begin{equation}
\|c\|^{2}=\sum_{ijkl}^{}c_{ijkl}^{2}=\sum_{\alpha\beta}^{}C_{\alpha\beta}^{2}.\label{eq:norm_el}\end{equation}

We remark here that Kelvin's notation is essential for the second
equality in both expressions \eqref{eq:scalar_el} and \eqref{eq:norm_el}.

The elasticity tensor's eigenvalues are positive real numbers due to
its symmetry and positive definiteness. We denote these eigenvalues
by $\lambda_{1},\lambda_{2},\ldots,\lambda_{r}$ and their
multiplicities by $m_{1},m_{2},\ldots,m_{r}$, where
$m_{1}+m_{2}+\cdots+m_{r}=6$. The norm of the elasticity tensor can
be expressed in terms of eigenvalues as\begin{equation}
\|c\|^{2}=m_{1}\lambda_{1}^{2} +
m_{2}\lambda_{2}^{2}+\cdots+m_{r}\lambda_{r}^{2}.\label{eq:norm_eig}\end{equation}

Scalar product \eqref{eq:scalar_el} and Frobenius norm
\eqref{eq:norm_el} are invariant under the action of orthogonal
group $O\left(3\right)$.

\section{Distance to isotropy\label{sec:Symmetry-classes}}

The harmonic decomposition of an elasticity tensor, which results
in its isotropic and anisotropic parts, has been used by several
researchers, notably Backus \cite{Backus}, Cowin \cite{Cowin} and
Baerheim \cite{Baerheim}, to study properties of this tensor.
Using this decomposition, we show that, for an arbitrary
elasticity tensor, the isotropic part is its closest isotropic
tensor, in the Euclidean sense. The norm of the anisotropic part,
which is the component orthogonal to the isotropic one, is the
distance between an arbitrary elasticity tensor and the class of
isotropic tensors, as discussed by Gazis et al. \cite{Gazis},
Fedorov \cite{Fedorov} and, recently, by Moakher and Norris
\cite{Moakher}. Combining these results, we discuss the distance
to isotropy using traces of the Voigt and dilatation tensors, the
Lamé parameters, as well as the two distinct eigenvalues of the
closest isotropic tensor. We show that these eigenvalues are the
effective isotropic parameters introduced by Voigt \cite{Voigt}
and discussed by Cowin \cite{Cowin}. Therefore, the distance of an
elasticity tensor to isotropy measures the deviation of its
eigenvalues from its effective isotropic parameters.

To derive and discuss the expression for distance, we consider an
elasticity tensor whose components with respect to a fixed
orthonormal basis, $\left\{ e_{1},e_{2},e_{3}\right\} $, are
$c_{ijkl}$. In terms of these components we consider the following
parameters introduced by Backus \cite{Backus} and Cowin \cite{Cowin}
for the harmonic decomposition of the elasticity tensor:
\begin{eqnarray}  \label{eq:lambda}  \lambda & = & \frac{1}{15} \left[c_{1111}+c_{2222}+c_{3333}+
4\left(c_{1122}+c_{1133}+c_{2233}\right) \right.\\
\nonumber & & \left.
-2\left(c_{1212}+c_{1313}+c_{2323}\right)\right], \\
\label{eq:mu} \mu & = &
\frac{1}{15}\left[c_{1111}+c_{2222}+c_{3333}-
\left(c_{1122}+c_{1133}+c_{2233}\right) \right.
\\ \nonumber & & \left. +3\left(c_{1212}+c_{1313}+c_{2323}\right)\right].\end{eqnarray}

The first parameter is related to the parameters obtained by
Moakher and Norris \cite{Moakher} by $\lambda=k-2\mu/3$; the
second one is the same as the parameter obtained by Moakher and
Norris \cite{Moakher}. Using definitions \eqref{eq:Voigt}, we
express $\lambda$ and $\mu$ in terms of traces of the Voigt and
dilatation tensors,

\begin{equation}
\lambda=\frac{1}{15}\left(2{\rm \mathrm{Tr}}D-{\rm
\mathrm{Tr}}V\right)\quad\mathrm{and}\quad\mu=\frac{1}{30}\left(3{\rm
\mathrm{Tr}}V-{\rm \mathrm{Tr}}D\right)\
,\label{eq:LambdaMu}\end{equation} which shows that these parameters
are invariants under the action of the orthogonal group $O(3)$.

According to Gazis et al. \cite[Corollary 2.1]{Gazis}, it follows
that tensor
\begin{equation} c^{iso}=\left[\begin{array}{cccccc}
\lambda+2\mu & \lambda & \lambda & 0 & 0 & 0\\
\lambda & \lambda+2\mu & \lambda & 0 & 0 & 0\\
\lambda & \lambda & \lambda+2\mu & 0 & 0 & 0\\
0 & 0 & 0 & 2\mu & 0 & 0\\
0 & 0 & 0 & 0 & 2\mu & 0\\
0 & 0 & 0 & 0 & 0 & 2\mu\end{array}\right]\label{eq:c_iso}\end{equation}

is positive-definite, and therefore one can view $\lambda$ and $\mu$
as the Lamé parameters of this isotropic tensor. Tensor $c^{iso}$
has two distinct eigenvalues, namely, \begin{equation}
\lambda_{1}^{iso}=3\lambda+2\mu=\frac{1}{3}{\rm
\mathrm{Tr}}D\label{eq:lambda1}\end{equation} and\begin{equation}
\lambda_{2}^{iso}=2\mu=\frac{1}{15}\left(3{\rm \mathrm{Tr}}V-{\rm
\mathrm{Tr}}D\right),\label{eq:lambda2}\end{equation} with
multiplicities $m_{1}=1$ and $m_{2}=5$. These two eigenvalues
coincide with the two effective isotropic moduli used by Voigt
\cite{Voigt} and Cowin \cite{Cowin}.

By a straightforward calculation we can show that
$\left(c-c^{iso}\right)\cdot c^{iso}=0$ and hence
$\left(c-c^{iso}\right)\perp c^{iso}$. In view of this
orthogonality, we denote
\begin{equation}
c_{\perp}^{iso}:=c-c^{iso},\label{eq:c_ortho}\end{equation} which
implies that \begin{equation}
\|c\|^{2}=\|c^{iso}\|^{2}+\|c_{\perp}^{iso}\|^{2}.\label{eq:HarmDecompIso}\end{equation}
Therefore, the square of the distance between an arbitrary
elasticity tensor and its orthogonal projection is \begin{equation}
d^{2}\left(c,
c^{iso}\right)=\|c_{\perp}^{iso}\|^{2}=\|c\|^{2}-\|c^{iso}\|^{2}.\label{eq:d_iso1}\end{equation}

From expressions \eqref{eq:lambda1} and \eqref{eq:lambda2} as well
as matrix \eqref{eq:c_iso}, it follows that the orthogonal
projection $c\mapsto c^{iso}$ is invariant under the action of the
orthogonal group $O(3)$. Therefore, no minimization is necessary
to obtain the distance to isotropy, and hence
\begin{equation}
d^2\left(c, {\mathcal Iso}\right)= d^{2}\left(c,
c^{iso}\right)=\|c_{\perp}^{iso}\|^{2}=\|c\|^{2}-\|c^{iso}\|^{2}.
\label{eq:d_isotropy}\end{equation}

It follows that tensor \eqref{eq:c_iso} is the \textit{closest
isotropic tensor} to a given elasticity tensor. Distance
\eqref{eq:d_isotropy} was obtained in terms of isotropic
parameters $k$ and $\mu$ by Moakher and Norris \cite{Moakher}.
Using expression \eqref{eq:c_iso}, we can write the distance
\eqref{eq:d_isotropy} in terms of parameters \eqref{eq:lambda} and
\eqref{eq:mu},
\begin{equation}
d^2\left(c, {\mathcal Iso}\right)=
\|c\|^{2}-3\left[\left(\lambda+2\mu\right)^{2}+2\left(\lambda^{2}+2\mu^{2}\right)\right].
\label{eq:d_iso2}\end{equation} Equivalently, using expressions
\eqref{eq:LambdaMu}, we can write distance \eqref{eq:d_isotropy} in
terms of traces of the Voigt and dilatation tensors,\begin{equation}
d^2\left(c, {\mathcal Iso}\right)=\|c\|^{2}-\frac{1}{15}\left(2{\rm
\mathrm{Tr}}^{2}D+3{\rm \mathrm{Tr}}^{2}V-2{\rm \mathrm{Tr}}V{\rm
\mathrm{Tr}}D\right).\label{eq:d_isoTrace}\end{equation} Also, using
expression \eqref{eq:norm_eig}, we can write this distance in terms
of the eigenvalues of the elasticity tensor and the two eigenvalues
of the closest isotropic tensor, which are expressions
\eqref{eq:lambda1} and \eqref{eq:lambda2}, to obtain
\begin{eqnarray} \label{eq:d_iso3}
d^2\left(c, {\mathcal Iso}\right) &= &
\sum_{\alpha=1}^{6}\lambda_{\alpha}^{2}-\left(\left(\lambda_{1}^{iso}\right)^{2}
+5\left(\lambda_{2}^{iso}\right)^{2}\right) \\
\nonumber & = &
\left(\lambda_{1}^{2}-\left(\lambda_{1}^{iso}\right)^{2}\right)+
\sum_{\alpha=2}^{6}\left(\lambda_{\alpha}^{2}-\left(\lambda_{2}^{iso}\right)^{2}\right).\end{eqnarray}

Examining expression \eqref{eq:d_iso3}, we see that an elasticity
tensor is close to being isotropic if and only if one of its eigenvalues
is close to $\lambda_{1}^{iso}$ and the other five are close to each
other and to $\lambda_{2}^{iso}$. In other words, the distance to
isotropy measures the deviation of the eigenvalues of the elasticity
tensor from the effective isotropic parameters.

\section{Distance to cubic symmetry\label{sec:Closeness-to-cubic}}

In this section, we discuss the distance of an elasticity tensor
to the class of cubic tensors. We fix ${\mathcal B}=\{e_1, e_2,
e_3\}$, which is an orthonormal basis of ${\mathbb R}^3$. For an
orthogonal transformation $A\in O(3)$, consider $A{\mathcal
B}:=\{Ae_1, Ae_2, Ae_3\}$, which is the corresponding orthonormal
basis, and can be identified with the orthogonal transformation
$A$.

We consider the components of elasticity tensor, $c_{ijkl}$, with
respect to orthonormal basis $A$. The following cubic tensor,
which has this orthonormal basis as its natural basis, has been
proposed by Moakher and Norris \cite{Moakher}:
\begin{equation} c^{cube}(A)=\left[\begin{array}{cccccc}
c_{1111}^{cube} & c_{1122}^{cube} & c_{1122}^{cube} & 0 & 0 & 0\\
c_{1122}^{cube} & c_{1111}^{cube} & c_{1122}^{cube} & 0 & 0 & 0\\
c_{1122}^{cube} & c_{1122}^{cube} & c_{1111}^{cube} & 0 & 0 & 0\\
0 & 0 & 0 & 2c_{1212}^{cube} & 0 & 0\\
0 & 0 & 0 & 0 & 2c_{1212}^{cube} & 0\\
0 & 0 & 0 & 0 & 0 & 2c_{1212}^{cube}\end{array}\right],\label{eq:c_cubic}\end{equation}
where\begin{equation}
c_{1111}^{cube}=\frac{1}{3}\left(c_{1111}+c_{2222}+c_{3333}\right),\label{eq:c_comp1111}\end{equation}
\begin{equation}
c_{1122}^{cube}=\frac{1}{3}\left(c_{1122}+c_{1133}+c_{2233}\right)\label{eq:c_comp1122}\end{equation}
and\begin{equation}
c_{1212}^{cube}=\frac{1}{3}\left(c_{1212}+c_{1313}+c_{2323}\right).\label{eq:c_comp1212}\end{equation}

According to Gazis et al. \cite[Corollary 2.1]{Gazis}, it follows
that tensor $c^{cube}(A)$ given by expression \eqref{eq:c_cubic}
is positive-definite and hence represents an elasticity tensor.
This elasticity tensor has three distinct
eigenvalues:\begin{equation}
\lambda_{1}^{cube}(A)=c_{1111}^{cube}+2c_{1122}^{cube},\label{eq:eigen_cubic1}\end{equation}
\begin{equation}
\lambda_{2}^{cube}(A)=c_{1111}^{cube}-c_{1122}^{cube}\label{eq:eigen_cubic2}\end{equation}
and\begin{equation}
\lambda_{3}^{cube}(A)=2c_{1212}^{cube},\label{eq:eigen_cubic3}\end{equation}
with multiplicities $m_{1}=1$, $m_{2}=2$ and $m_{3}=3$, as
expected in view of the coordinate-free characterization
formulated by Bóna et al. \cite{BBS1}. As we will see later,
eigenvalue \eqref{eq:eigen_cubic1} is an orthogonal invariant. The
other two eigenvalues, \eqref{eq:eigen_cubic2} and
\eqref{eq:eigen_cubic3}, are not invariant under rotations of $c$,
and hence depend on orientation $A$. These eigenvalues were
considered also by Moakher \cite{Moakher06} and coincide with
parameters $a$, $c$ and $b$, respectively, obtained by Moakher and
Norris \cite{Moakher}.

Using matrices \eqref{eq:Chapman} and \eqref{eq:c_cubic} as well as
expressions \eqref{eq:c_comp1111} -- \eqref{eq:c_comp1212}, one can
show by direct calculations that $\left(c-c^{cube}(A)\right)\cdot
c^{cube}(A)=0$, which justifies the notation
$c_{\perp}^{cube}(A):=c-c^{cube}(A)$, and implies
that\begin{equation}
\|c\|^{2}=\|c^{cube}(A)\|^{2}+\|c_{\perp}^{cube}(A)\|^{2}.\label{eq:HarmDecompCube}\end{equation}
Therefore, the square of the distance between a given elasticity
tensor, $c$, and its orthogonal projection, $c^{cube}(A)$, is
\begin{equation} d^{2}\left(c,
c^{cube}(A)\right)=\|c_{\perp}^{cube}(A)\|^{2}=\|c\|^{2}-\|c^{cube}(A)\|^{2}.
\label{eq:d_cubic1}\end{equation} This expression was obtained in
terms of the cubic parameters $a$, $b$ and $c$ by Moakher and Norris
\cite{Moakher}. Using expression \eqref{eq:norm_eig}, we can write
$d^{2}\left(c, c^{cube}(A)\right)$ in terms of the eigenvalues of
the given elasticity tensor and eigenvalues \eqref{eq:eigen_cubic1},
\eqref{eq:eigen_cubic2} and \eqref{eq:eigen_cubic3}. Hence, we write
\begin{multline} d^{2}\left(c, c^{cube}(A)\right)
=\sum_{\alpha=1}^{6}
\lambda_{\alpha}^{2}-\left(\left(\lambda_{1}^{cube}(A)\right)^{2}+
2\left(\lambda_{2}^{cube}(A)\right)^{2} + 3\left(\lambda_{3}^{cube}(A)\right)^{2}\right)\\
=\left(\lambda_{1}^{2}-\left(\lambda_{1}^{cube}(A)\right)^{2}\right)
+\sum_{\alpha=2}^{3}\left(\lambda_{\alpha}^{2}-\left(\lambda_{2}^{cube}(A)\right)^{2}\right)
+\sum_{\alpha=4}^{6}\left(\lambda_{\alpha}^{2}-\left(\lambda_{3}^{cube}(A)\right)^{2}\right),
\label{eq:d_cubic2}\end{multline} which expresses the deviations
of the eigenvalues of the elasticity tensor from its orthogonal
projection.

In order to obtain distance $d\left(c, {\mathcal Cube}\right)$ we
have to minimize formula \eqref{eq:d_cubic2} for $A\in O(3)$,
which is equivalent to maximizing function
\begin{eqnarray}
\nonumber f\left(\lambda_{1}^{cube}(A), \lambda_{1}^{cube}(A),
\lambda_{1}^{cube}(A)\right)= \\
\left(\lambda_{1}^{cube}(A)\right)^{2}+
2\left(\lambda_{2}^{cube}(A)\right)^{2} +
3\left(\lambda_{3}^{cube}(A)\right)^{2}. \label{fcube}
\end{eqnarray}

We now prove that $\lambda_{1}^{cube}(A)$ is an
orthogonal invariant and we can find an orthogonal invariant
relation between $\lambda_{2}^{cube}(A)$ and
$\lambda_{3}^{cube}(A)$. Therefore, the process of maximizing
function \eqref{fcube} can be simplified.

First, we notice that the two linear invariants associated with
the elasticity tensor, ${\rm \mathrm{Tr}}V$ and ${\rm
\mathrm{Tr}}D$, remain invariant under orthogonal projection
$c\mapsto c^{cube}$. Let us denote the Voigt and dilatation
tensors associated with $c$ and its orthogonal projection,
$c^{cube}(A)$, by $V$, $V^{cube}(A)$ and $D$, $D^{cube}(A)$,
respectively. In view of expressions \eqref{eq:Voigt} and
\eqref{eq:c_comp1111} -- \eqref{eq:c_comp1212}, it follows
that\begin{equation} \operatorname{Tr}V =
\operatorname{Tr}V^{cube}(A) \quad \mathrm{and}
\quad\operatorname{Tr}D =\operatorname{Tr}D^{cube}(A),
\label{eq:invariant_tr_cub} \forall A\in O(3). \end{equation}

Examining expressions \eqref{eq:lambda1} and
\eqref{eq:eigen_cubic1}, or the invariance of $\operatorname{Tr}D$
under both projections $c\mapsto c^{cube}$ and $c\mapsto c^{iso}$,
we see that \begin{equation}
\lambda_{1}^{cube}(A)=\lambda_{1}^{iso}=\frac{1}{3}\operatorname{Tr}D,
\label{eq:l1c_inv}\end{equation} which shows that isotropy and
cubic symmetry share one effective elasticity parameter; it is an
isotropic invariant. In other words, $\lambda_{1}^{cube}$ remains
unchanged if the elasticity tensor is subject to an orthogonal
transformation or its components are expressed in a different
orthonormal basis. The other two eigenvalues of the cubic tensor,
$\lambda_{2}^{cube}(A)$ and $\lambda_{3}^{cube}(A)$, are not
isotropic invariants. However, examining expressions
\eqref{eq:eigen_cubic2}, \eqref{eq:eigen_cubic3} and
\eqref{eq:lambda2}, we see that their combination is an isotropic
invariant: \begin{equation}
2\lambda_{2}^{cube}(A)+3\lambda_{3}^{cube}(A)=5\lambda_{2}^{iso}
=\operatorname{Tr}V - \frac{1}{3}\operatorname{Tr}D.
\label{eq:l23c_inv}\end{equation}

Digressing, we note that, physically, the split of
$\lambda_{2}^{iso}$, which has multiplicity $m_{2}^{iso}=5$, into
$\lambda_{2}^{cube}$ and $\lambda_{3}^{cube}$, which have
multiplicities $m_{2}^{cube}=2$ and $m_{3}^{cube}=3$, is related to
the distinction between the eigenvalues that govern pure and simple
shears, respectively, for cubic continua.

Expression \eqref{eq:l23c_inv} is a consequence of the invariance
for $\operatorname{Tr}V$ under both projections $c\mapsto c^{cube}$
and $c\mapsto c^{iso}$, which can be written as
\begin{eqnarray}
\label{eq:inv_tr_v} \operatorname{Tr}V^{cube}(A) & = &
\lambda_{1}^{cube}(A)+2\lambda_{2}^{cube}(A)+3\lambda_{3}^{cube}(A)
\\
\nonumber = \operatorname{Tr}V^{iso} & = &
\lambda_{1}^{iso}+5\lambda_{2}^{iso} = \operatorname{Tr}V, \
\forall A\in O(3). \end{eqnarray} In order to obtain the distance
to the subspace of cubic tensors, we must minimize the value of
expression \eqref{eq:d_cubic2} or maximize the value of
expression \eqref{fcube}, using \eqref{eq:l1c_inv} and
\eqref{eq:l23c_inv}. Since this minimization is performed over the
compact group $O(3)$, it follows that the minimum of expression
\eqref{eq:d_cubic2} can be reached for an orthogonal
transformation, $A_e$. Hence, there exists $A_e\in O(3)$, such
that
\begin{equation}
d\left(c, {\mathcal Cube}\right)=d\left(c, c^{cube}(A_e)\right).
\label{cubic_de} \end{equation} Therefore, $c^{cube}(A_e)$ is the
\textit{closest cubic tensor} to elasticity tensor $c$.
Eigenvalues $\lambda_1^{cube}(A_e)$, $\lambda_2^{cube}(A_e)$,
$\lambda_3^{cube}(A_e)$ represent the \textit{effective cubic
parameters}, while $A_e$ is the \textit{effective cubic
orientation} for the elasticity tensor $c$.

Since changes of orthonormal basis in $\mathbb{R}^{3}$ affect
neither the six eigenvalues of the elasticity tensor nor the
isotropic invariant, $\lambda_{1}^{cube}$, using expression
\eqref{eq:l23c_inv}, we state expression \eqref{eq:d_cubic2} as
\begin{eqnarray} \label{eq:dc_diso} d^{2}\left(c, c^{cube}(A)\right)
& =& d^{2}\left(c, {\mathcal Iso}\right)
-\frac{10}{3}\left(\lambda_{2}^{iso}-\lambda_{2}^{cube}(A)\right)^{2}
\\ \nonumber &= & d^{2}\left(c, {\mathcal Iso}\right)
-\frac{15}{2}\left(\lambda_{2}^{iso}-\lambda_{3}^{cube}(A)\right)^{2}.\end{eqnarray}

Both expressions
$10\left(\lambda_{2}^{iso}-\lambda_{2}^{cube}(A)\right)^{2}/3$ and
$15\left(\lambda_{2}^{iso}-\lambda_{3}^{cube}(A)\right)^{2}/2$
represent the square of the distance between $c^{cube}(A)$ and the
class of isotropic tensors. In other words, we can rewrite
expression \eqref{eq:dc_diso} as\begin{equation} d^{2}\left(c,
c^{cube}(A)\right)=d^{2}\left(c, {\mathcal Iso}\right)-
d^{2}\left(c^{cube}(A), {\mathcal
Iso}\right).\label{eq:dc_diso2}\end{equation} Expression
\eqref{eq:dc_diso2} was implemented in the same context by Moakher
and Norris \cite{Moakher}.

If in expression \eqref{eq:dc_diso2} we choose $A$ to be the
effective cubic orientation $A_e$ and use equation
\eqref{cubic_de}, we obtain
\begin{equation} d^{2}\left(c,
{\mathcal Cube}\right)=d^{2}\left(c, {\mathcal Iso}\right)-
d^{2}\left(c^{cube}(A_e), {\mathcal
Iso}\right),\label{eq:dc_diso3}\end{equation} which illustrates
the fact that, for a given elasticity tensor, its distance to
isotropy is greater than its distance to any other
material-symmetry class.

If in expression \eqref{eq:d_cubic2} we choose $A$ to be the
effective cubic orientation $A_e$ and use equation
\eqref{cubic_de}, we obtain
\begin{multline} d^{2}\left(c, {\mathcal Cube}\right)
=\sum_{\alpha=1}^{6}
\lambda_{\alpha}^{2}-\left(\left(\lambda_{1}^{cube}(A_e)\right)^{2}+
2\left(\lambda_{2}^{cube}(A_e)\right)^{2} + 3\left(\lambda_{3}^{cube}(A_e)\right)^{2}\right)\\
=\left(\lambda_{1}^{2}-\left(\lambda_{1}^{cube}(A_e)\right)^{2}\right)
+\sum_{\alpha=2}^{3}\left(\lambda_{\alpha}^{2}-\left(\lambda_{2}^{cube}(A_e)\right)^{2}\right)
+\sum_{\alpha=4}^{6}\left(\lambda_{\alpha}^{2}-\left(\lambda_{3}^{cube}(A_e)\right)^{2}\right),
\label{eq:d_cubicparam}\end{multline} which expresses the
deviations of the elasticity tensor from cubic symmetry, in terms
of the effective cubic parameters. In other words, an elasticity
tensor is close to having cubic symmetry if one of its eigenvalues
is close to $\lambda_{1}^{cube}$, two of them are close to one
another and to $\lambda_{2}^{cube}(A_e)$, and the other three are
close to each other and to $\lambda_{3}^{cube}(A_e)$.

\section{Distance to transverse isotropy\label{sec:Closeness-to-transverse}}

In this section, we discuss the distance of an elasticity tensor
to the class of transversely isotropic tensors. We consider the
components of elasticity tensor, $c_{ijkl}$, with respect to
orthonormal basis $A$. The following transversely isotropic
tensor, which has this orthonormal basis as its natural basis, has
been proposed by Moakher and Norris \cite{Moakher}:
\begin{equation} c^{TI}(A)=\left[\begin{array}{cccccc}
c_{1111}^{TI} & c_{1122}^{TI} & c_{1133}^{TI} & 0 & 0 & 0\\
c_{1122}^{TI} & c_{1111}^{TI} & c_{1133}^{TI} & 0 & 0 & 0\\
c_{1133}^{TI} & c_{1133}^{TI} & c_{3333}^{TI} & 0 & 0 & 0\\
0 & 0 & 0 & 2c_{2323}^{TI} & 0 & 0\\
0 & 0 & 0 & 0 & 2c_{2323}^{TI} & 0\\
0 & 0 & 0 & 0 & 0 & c_{1111}^{TI}-c_{1122}^{TI}\end{array}\right],\label{eq:c_ti}\end{equation}
where\begin{equation}
c_{1111}^{TI}=\frac{1}{8}\left(3c_{1111}+3c_{2222}+2c_{1122}+4c_{1212}\right),\label{eq:ti_comp1111}\end{equation}
\begin{equation}
c_{1122}^{TI}=\frac{1}{8}\left(c_{1111}+c_{2222}+6c_{1122}-4c_{1212}\right),\label{eq:ti_comp1122}\end{equation}
\begin{equation}
c_{1133}^{TI}=\frac{1}{2}\left(c_{1133}+c_{2233}\right),\label{eq:ti_comp1133}\end{equation}
\begin{equation}
c_{3333}^{TI}=c_{3333}\label{eq:ti_comp3333}\end{equation}
and\begin{equation}
c_{2323}^{TI}=\frac{1}{2}\left(c_{2323}+c_{1313}\right).\label{eq:ti_comp2323}\end{equation}

According to Gazis et al. \cite[Corollary 2.1]{Gazis}, it follows
that tensor $c^{TI}(A)$ given by expression \eqref{eq:c_ti} is
positive-definite and hence it represents an elasticity tensor.
This elasticity tensor has four distinct
eigenvalues:\begin{equation}
\lambda_{1}^{TI}(A)=\frac{c_{1111}^{TI}+c_{1122}^{TI}+c_{3333}^{TI}+
\sqrt{\left(c_{1111}^{TI}+c_{1122}^{TI}-c_{3333}^{TI}\right)^{2}+
8\left(c_{1133}^{TI}\right)^{2}}}{2},\label{eq:eigen_ti1}\end{equation}
\begin{equation}
\lambda_{2}^{TI}(A)=\frac{c_{1111}^{TI}+c_{1122}^{TI}+c_{3333}^{TI}-
\sqrt{\left(c_{1111}^{TI}+c_{1122}^{TI}-c_{3333}^{TI}\right)^{2}+
8\left(c_{1133}^{TI}\right)^{2}}}{2},\label{eq:eigen_ti2}\end{equation}
\begin{equation}
\lambda_{3}^{TI}(A)=c_{1111}^{TI}-c_{1122}^{TI}\label{eq:eigen_ti3}\end{equation}
and\begin{equation}
\lambda_{4}^{TI}(A)=2c_{2323}^{TI},\label{eq:eigen_ti4}\end{equation}
with multiplicities $m_{1}=1$, $m_{2}=1$, $m_{3}=2$ and $m_{4}=2$,
as expected in view of Bóna et al. \cite{BBS1}. Again, unlike
eigenvalues \eqref{eq:lambda1} and \eqref{eq:lambda2}, these
eigenvalues are not invariant under rotations of $c$, and hence
they explicitly depend on orthonormal basis $A$. These eigenvalues
were also considered by Moakher \cite{Moakher06}. The relations
between these eigenvalues and the parameters obtained by Moakher
and Norris \cite{Moakher} are\begin{equation}
\lambda_{1,2}^{TI}(A)=\frac{a+b\pm\sqrt{\left(a-b\right)^{2}+c^{2}}}{2},\
\lambda_{3}^{TI}(A)=f,\
\lambda_{4}^{TI}(A)=g.\label{eq:abcfg}\end{equation}

Using matrices \eqref{eq:Chapman} and \eqref{eq:c_ti} as well as
expressions \eqref{eq:ti_comp1111} -- \eqref{eq:ti_comp2323}, one
can show by direct calculations that $\left(c-c^{TI}(A)\right)\cdot
c^{TI}(A)=0,$ which justifies the notation
$c_{\perp}^{TI}(A):=c-c^{TI}(A)$, and implies that \[
\|c\|^{2}=\|c^{TI}(A)\|^{2}+\|c_{\perp}^{TI}(A)\|^{2}.\] Therefore
the square of the distance between a given elasticity tensor, $c$,
and its orthogonal projection, $c^{TI}(A)$, is\begin{equation}
d^{2}\left(c,c^{TI}(A)\right)=\|c_{\perp}^{TI}(A)\|^{2}=\|c\|^{2}-\|c^{TI}(A)\|^{2}.
\label{eq:d_ti1}\end{equation} Expression \eqref{eq:d_ti1} was
obtained by Moakher and Norris \cite{Moakher} using five parameters:
$a$, $b$, $c$, $f$ and $g$. Using expression \eqref{eq:norm_eig}, we
can write $d^{2}\left(c,c^{TI}(A)\right)$ in terms of the
eigenvalues of the given elasticity tensor and the eigenvalues
\eqref{eq:eigen_ti1} -- \eqref{eq:eigen_ti4} of the orthogonal
projection. Hence, we write
\begin{multline}
d^{2}\left(c,c^{TI}(A)\right)=\sum_{\alpha=1}^{6}\lambda_{\alpha}^{2}-
\left(\left(\lambda_{1}^{TI}(A)\right)^{2}+\left(\lambda_{2}^{TI}(A)\right)^{2}+
2\left(\lambda_{3}^{TI}(A)\right)^{2}+2\left(\lambda_{4}^{TI}(A)\right)^{2}\right)\label{eq:d_ti2}\\
=\left(\lambda_{1}^{2}-\left(\lambda_{1}^{TI}(A)\right)^{2}\right)+
\left(\lambda_{2}^{2}-\left(\lambda_{2}^{TI}(A)\right)^{2}\right)+
\sum_{\alpha=3}^{4}\left(\lambda_{\alpha}^{2}-\left(\lambda_{3}^{TI}(A)\right)^{2}\right)+
\sum_{\alpha=5}^{6}\left(\lambda_{\alpha}^{2}-\left(\lambda_{4}^{TI}(A)\right)^{2}\right),\end{multline}
which expresses the deviations of the eigenvalues of the
elasticity tensor from those of the corresponding orthogonal
projection.

Using expression \eqref{eq:d_iso3}, we write expression
\eqref{eq:d_ti2} as
\begin{eqnarray}
\label{eq:d_ti3} d^{2}\left(c,c^{TI}(A)\right)  = d^{2}\left(c,
{\mathcal Iso}\right) + \left(\lambda_{1}^{iso}\right)^{2} +
5\left(\lambda_{2}^{iso}\right)^{2} \\
\nonumber -\left(\left(\lambda_{1}^{TI}(A)\right)^{2} +
\left(\lambda_{2}^{TI}(A)\right)^{2} +
2\left(\lambda_{3}^{TI}(A)\right)^{2} +
2\left(\lambda_{4}^{TI}(A)\right)^{2}\right).\end{eqnarray}

In order to obtain distance $d\left(c, {\mathcal TI}\right)$, we
have to minimize the above formula \eqref{eq:d_ti3}, for $A$ in
$O(3)$, which is equivalent to maximizing function
\begin{eqnarray}
\label{fti} f\left(\lambda_{1}^{TI}(A), \lambda_{2}^{TI}(A),
\lambda_{3}^{TI}(A), \lambda_{4}^{TI}(A)\right)= \\
\nonumber \left(\lambda_{1}^{TI}(A)\right)^{2} +
\left(\lambda_{2}^{TI}(A)\right)^{2} +
2\left(\lambda_{3}^{TI}(A)\right)^{2} +
2\left(\lambda_{4}^{TI}(A)\right)^{2}.
\end{eqnarray}

Even though these expressions are not invariant under orthogonal
transformations, the two linear invariants associated with the
elasticity tensor, $\operatorname{Tr}V$ and $\operatorname{Tr}D$,
remain invariant. We denote the Voigt and dilatation tensors
associated with $c$ and its projection, $c^{TI}(A)$, by $V$, $
V^{TI}(A)$ and $D$, $D^{TI}(A)$, respectively. In view of
expressions \eqref{eq:Voigt} and \eqref{eq:ti_comp1111} --
\eqref{eq:ti_comp2323}, it follows that \begin{equation}
\operatorname{Tr}V=\operatorname{Tr}V^{TI}(A) \quad \mathrm{and}
\quad \operatorname{Tr}D = \operatorname{Tr}D^{TI}(A), \ \forall
A\in O(3), \label{eq:invariant_tr_ti}\end{equation} which are
analogous to expressions \eqref{eq:invariant_tr_cub}.

Examining expressions \eqref{eq:eigen_ti1} --
\eqref{eq:eigen_ti4}, or the invariance of $\operatorname{Tr}V$
under projections $c\mapsto c^{TI}$ and $c\mapsto c^{iso}$, we
obtain
\begin{eqnarray} \label{eq:inv_v_ti} \operatorname{Tr}V^{TI}(A) & = & \lambda_{1}^{TI}(A)+
\lambda_{2}^{TI}(A) + 2\lambda_{3}^{TI}(A) + 2\lambda_{4}^{TI}(A)
\\ \nonumber =
\operatorname{Tr}V^{iso} & =&
\lambda_{1}^{iso}+5\lambda_{2}^{iso}=\operatorname{Tr}V, \ \forall
A\in O(3).
\end{eqnarray}

We note that $\operatorname{Tr}D^{TI}$ can be expressed in terms of
two eigenvalues $\lambda_{1}^{TI},\ \lambda_{2}^{TI}$ and another
orthogonal invariant, which is used by Bóna et al. \cite{BBS1} for
characterizing transversely isotropic symmetry but does not appear
explicitly in the distance function; it is referred to by Bóna et
al. \cite{BBS1} as $\gamma$. Therefore, the invariance of
$\operatorname{Tr}D^{TI}$ does not have an influence on minimizing
expression \eqref{eq:d_ti2}.

To obtain the distance to the class of transversely isotropic
tensors, we minimize expression \eqref{eq:d_ti2} under the
restriction given by expression \eqref{eq:inv_v_ti}. This is
equivalent to maximizing function \eqref{fti} under the same
restriction. Using expression \eqref{eq:inv_v_ti}, we write
expression \eqref{eq:d_ti3} as
\begin{multline} d^{2}\left(c, c^{TI}(A)\right) = d^{2}\left(c,
{\mathcal Iso}\right) +
\left(\lambda_{1}^{iso}\right)^{2}+5\left(\lambda_{2}^{iso}\right)^{2}\\
-\left[\left(\lambda_{1}^{TI}(A)\right)^{2} +
\left(\lambda_{2}^{TI}(A)\right)^{2}+2\left(\lambda_{3}^{TI}(A)\right)^{2}\right.
\\
\left. +\frac{1}{2}\left(\lambda_{1}^{iso}+5\lambda_{2}^{iso}-
\left(\lambda_{1}^{TI}(A)+\lambda_{2}^{TI}(A)+2\lambda_{3}^{TI}(A)\right)\right)^{2}\right],
\label{eq:d_ti4}\end{multline} which expresses the square of the
distance to transversely isotropic projection.

The minimum value of \eqref{eq:d_ti4} represents the square of the
distance of the given tensor, $c$, to transverse isotropy. Since
the minimization is performed over the compact group $O(3)$, it
follows that the minimum can be reached for an orthogonal
transformation, $A_e$. Hence, there exists $A_e\in O(3)$, such
that
\begin{equation} d\left(c, {\mathcal TI}\right)=d\left(c,
c^{TI}(A_e)\right). \label{effective_dti}
\end{equation}

Therefore, $c^{TI}(A_e)$ represents the \textit{closest isotropic
tensor} to elasticity tensor $c$. Eigenvalues
$\lambda_1^{TI}(A_e)$, $\lambda_2^{TI}(A_e)$,
$\lambda_3^{TI}(A_e)$, $\lambda_4^{TI}(A_e)$ represent the
\textit{effective transverse isotropic parameters}, while
orthonormal basis $A_e$ gives the \textit{effective orientation}
of the closest transversely isotropic counterpart to $c$.

If in expression \eqref{eq:d_ti3} we choose $A$ to be the
effective transversely isotropic orientation, $A_e$, and use
equation \eqref{effective_dti}, we obtain
\begin{equation}
d^2\left(c, {\mathcal TI}\right)=d^2\left(c, {\mathcal Iso}\right) -
d^2\left(c^{TI}(A_e), {\mathcal Iso}\right). \label{dti_effective}
\end{equation}

If in expression \eqref{eq:d_ti2} we choose $A$ to be the
effective transversely isotropic orientation, $A_e$, and use
equation \eqref{effective_dti}, we obtain
\begin{multline}
d^{2}\left(c, {\mathcal TI}\right) =
\sum_{\alpha=1}^{6}\lambda_{\alpha}^{2}-
\left(\left(\lambda_{1}^{TI}(A_e)\right)^{2}+\left(\lambda_{2}^{TI}(A_e)\right)^{2}+
2\left(\lambda_{3}^{TI}(A_e)\right)^{2}+2\left(\lambda_{4}^{TI}(A_e)\right)^{2}\right)\label{eq:d_tief}\\
=\left(\lambda_{1}^{2}-\left(\lambda_{1}^{TI}(A_e)\right)^{2}\right)+
\left(\lambda_{2}^{2}-\left(\lambda_{2}^{TI}(A_e)\right)^{2}\right)+
\sum_{\alpha=3}^{4}\left(\lambda_{\alpha}^{2}-\left(\lambda_{3}^{TI}(A_e)\right)^{2}\right)+
\sum_{\alpha=5}^{6}\left(\lambda_{\alpha}^{2}-\left(\lambda_{4}^{TI}(A_e)\right)^{2}\right),\end{multline}
which expresses the deviations of the elasticity tensor from
transverse isotropy, in terms of the effective transversely
isotropic parameters. In other words, an elasticity tensor is
close to having transversely isotropic symmetry if one of its
eigenvalues is close to $\lambda_{1}^{TI}(A_e)$, one is close to
$\lambda_{2}^{TI}(A_e)$, two of them are close to one another and
close to $\lambda_{3}^{TI}(A_e)$, and the other two are close to
one another and close to $\lambda_{4}^{TI}(A_e)$.

\section{Examples}

In this section, we use the expressions derived above to illustrate
the finding of the distance between a given elasticity tensor and
isotropy, cubic symmetry and transverse isotropy. We consider both
strongly and weakly anisotropic materials. In both cases, we present
a density-scaled elasticity tensor in the Kelvin's notation stated
in expression \eqref{eq:Chapman}.

\subsection{Strong anisotropy}

Let us consider a strongly anisotropic case, namely, \begin{equation}
\left[\begin{array}{cccccc}
56.60 & 8.98 & 3.45 & 0 & 0 & 0\\
8.98 & 56.60 & 3.45 & 0 & 0 & 0\\
3.45 & 3.45 & 16.43 & 0 & 0 & 0\\
0 & 0 & 0 & 3.60 & 0 & 0\\
0 & 0 & 0 & 0 & 3.60 & 0\\
0 & 0 & 0 & 0 & 0 &
47.62\end{array}\right]\left[\mathrm{\frac{km^{2}}{s^{2}}}\right],
\label{eq:Biotite}\end{equation} which corresponds to biotite and
exhibits transverse isotropy, as quoted by Thomsen \cite{Thomsen}.
Its anisotropy parameters defined by Thomsen \cite{Thomsen} are
$\gamma=6.12$, $\delta=-0.39$, $\varepsilon=1.22$; their magnitude
indicates strong anisotropy. The eigenvalues of this tensor are
$66.07$ and $15.95$ with multiplicity one, and $47.62$ and $3.60$
with multiplicity two. Expression \eqref{eq:Biotite} is considered
with respect to a natural basis of the transversely isotropic
medium.

\subsubsection{Relation to isotropy}

Using equations \eqref{eq:lambda1} and \eqref{eq:lambda2}, we obtain
the eigenvalues of the closest isotropic tensor,
$\lambda_{1}^{iso}=53.80$ and
$\lambda_{2}^{iso}=\ldots=\lambda_{6}^{iso}=26.13$, which are the
effective isotropic parameters, as well as $\lambda=9.22$ and
$\mu=13.06$, which are the Lamé parameters of the closest
representative of the isotropic subspace. Using these values and any
of expressions \eqref{eq:d_iso2} -- \eqref{eq:d_iso3}, we obtain the
distance to isotropy: $d\left(c, {\mathcal Iso}\right)=53.59\
\mathrm{km^{2}/s^{2}}$.

\subsubsection{Relation to cubic symmetry\label{sub:Relation-to-cubic}}

Using expressions \eqref{eq:eigen_cubic1}, \eqref{eq:eigen_cubic2}
and \eqref{eq:eigen_cubic3}, we obtain the eigenvalues of the
corresponding cubic tensor, with respect to the fixed natural basis
of elasticity tensor $c$: $\lambda_{1}^{cube}=53.80$,
$\lambda_{2}^{cube}=37.92$ and $\lambda_{3}^{cube}=18.27$. For an
arbitrary orthonormal basis $A$, consider $c^{cube}(A)$ the cubic
orthogonal projection. Invoking one of the equalities in expression
\eqref{eq:dc_diso} and using $d\left(c, {\mathcal Iso}\right)$ and
$\lambda_{2}^{iso}$ from above, we write\begin{eqnarray}
\label{eq:AGU1} d^{2}\left(c, c^{cube}(A)\right) & =&  d^{2}\left(c,
{\mathcal Iso}\right) -
\frac{10}{3}\left(\lambda_{2}^{iso}-\lambda_{2}^{cube}(A)\right)^{2}
\\ \nonumber &= & 53.59^{2}- \frac{10}{3}
\left(26.13-\lambda_{2}^{cube}(A)\right)^{2}.
\end{eqnarray}

To find the distance to the cubic-symmetry subspace, we look for
$\lambda_{2}^{cube}(A)$ that minimizes this expression. The
resulting distance $d\left(c, {\mathcal Cube}\right)=49.07\
\mathrm{km^{2}/s^{2}}$ is achieved by rotation of the natural basis
of elasticity tensor $c$ with the effective transformation
\[
A_e=\left[\begin{array}{ccr}
0 & 0 & 1\\
\\-0.82 & -0.58 & 0\\
\\0.58 & -0.82 & 0\end{array}\right].\]
The resulting cubic orthogonal projection $c^{cube}(A_e)$ represents
the closest cubic tensor to $c$. Its eigenvalues, which are the
effective cubic parameters for elasticity tensor \eqref{eq:Biotite},
are $\lambda_{1}^{cube}(A_e)=53.80$, $\lambda_{2}^{cube}(A_e)=12.64$
and $\lambda_{3}^{cube}(A_e)=18.27$. Minimizations described herein
lend themselves to such software as Maple and Mathematica.

\subsubsection{Relation to transverse isotropy\label{sub:Relation-to-transverse}}

Minimizing expression \eqref{eq:d_ti4} subject to
$\lambda_{1}^{TI}$, $\lambda_{2}^{TI}$, $\lambda_{3}^{TI}$ and
$\lambda_{4}^{TI}$ being positive, we obtain $d\left(c, {\mathcal
TI}\right)=0$, as expected. Since tensor \eqref{eq:Biotite} is
expressed in natural coordinates, we can obtain the same result
using expressions \eqref{eq:eigen_ti1} -- \eqref{eq:eigen_ti4} to
calculate the eigenvalues of the effective transversely isotropic
tensor, and verify that they are the same as the eigenvalues of
tensor \eqref{eq:Biotite}.

\subsection{Weak anisotropy}

Let us consider a weakly anisotropic case, namely,\begin{equation}
\left[\begin{array}{cccccc}
4.00 & 2.06 & 2.10 & -0.07 & 0.01 & -0.03\\
2.06 & 3.83 & 1.96 & 0.17 & -0.07 & 0.18\\
2.10 & 1.96 & 3.96 & 0.16 & 0.04 & -0.13\\
-0.07 & 0.17 & 0.16 & 2.00 & 0.22 & -0.14\\
0.01 & -0.07 & 0.04 & 0.22 & 1.76 & 0.02\\
-0.03 & 0.18 & -0.13 & -0.14 & 0.02 & 2.22\end{array}\right]
\left[\mathrm{\frac{km^{2}}{s^{2}}}\right],\label{eq:TenVlad}\end{equation}
which are the density-scaled elasticity parameters used by Dewangan
and Grechka \cite{Dewangan}. This is a generally anisotropic medium:
it does not exhibit any material symmetry. The eigenvalues are
distinct from each other: $8.02$, $2.39$, $2.16$, $1.86$, $1.82$,
$1.52$.

\subsubsection{Relation to isotropy}

The eigenvalues of the closest isotropic tensor are $\lambda_{1}^{iso}=8.01$
and $\lambda_{2}^{iso}=\ldots=\lambda_{6}^{iso}=1.95$, which are
the effective isotropic parameters, and the Lamé parameters are $\lambda=2.02$
and $\mu=0.98$. Herein, the anisotropy is very weak: the distance
to isotropy is $0.724\ \mathrm{km^{2}/s^{2}}$. Notably, by inspection
of the corresponding entries in matrices \eqref{eq:c_iso} and \eqref{eq:TenVlad}
or by comparison of the eigenvalues of $c$ and $c^{iso}$, we can
expect the closeness of this tensor to isotropy.

\subsubsection{Relation to cubic symmetry}

Proceeding in a manner analogous to the one described in Section
\ref{sub:Relation-to-cubic}, we obtain $d\left(c, {\mathcal
Cube}\right)=0.64\ \mathrm{km^{2}/s^{2}}$, where the effective
orientation of the closest cubic tensor to elasticity tensor
\eqref{eq:TenVlad} is \[ A_e=\left[\begin{array}{ccr}
0.77 & 0.36 & 0.53\\
\\-0.60 & 0.13 & 0.79\\
\\0.21 & -0.93 & 0.31\end{array}\right].\]
The effective cubic parameters of the elasticity tensor
\eqref{eq:TenVlad} are $\lambda_{1}^{cube}(A_e)=8.01$,
$\lambda_{2}^{cube}(A_e)=0.59$ and $\lambda_{3}^{cube}(A_e)=2.08$.

\subsubsection{Relation to transverse isotropy}

Proceeding in a manner analogous to the one described in Section
\ref{sub:Relation-to-transverse}, we obtain $d\left(c, {\mathcal
TI}\right)=0.57\ \mathrm{km^{2}/s^{2}}$, where the effective
orientation of the closest transversely isotropic tensor to
elasticity tensor \eqref{eq:TenVlad} is \[
A_e=\left[\begin{array}{ccr}
0.76 & 0.53 & 0.38\\
\\-0.58 & 0.29 & 0.76\\
\\0.29 & -0.80 & 0.53\end{array}\right].\]

The effective transversely isotropic parameters of the elasticity
tensor \eqref{eq:TenVlad} are $\lambda_{1}^{TI}(A_e)=8.01$,
$\lambda_{2}^{TI}(A_e)=1.62$, $\lambda_{3}^{TI}(A_e)=2.05$ and
$\lambda_{4}^{TI}(A_e)=2.02$.

\begin{acknowledgement*}
The authors wish to acknowledge fruitful discussions with Andrej
Bóna, Ça\u{g}r\i~Diner and Michael G.~Rochester, and the editorial
work of David Dalton and Leslie McNab. The research was done in
the context of The Geomechanics Project. The research of I.B. was
supported also by Grant PN II IDEI 398 of Romanian Ministry of
Education. The research of M.A.S. was supported also by NSERC.
\end{acknowledgement*}

\end{document}